\documentclass{pasj00}

\draft

\begin{document}
\SetRunningHead{S. Kato}{}
\Received{2012/00/00}
\Accepted{2012/00/00}

\title{A Resonantly-Excited Disk-Oscillation Model of High-Frequency QPOs of Microquasars} 

\author{Shoji \textsc{Kato}}
\affil{2-2-2 Shikanoda-Nishi, Ikoma-shi, Nara, 630-0114}
\email{kato@gmail.com, kato@kusastro.kyoto-u.ac.jp}

%


\maketitle

\begin{abstract} 
A possible model of twin high-frequency QPOs (HF QPOs) of microquasars is examined.
The disk is assumed to have global magnetic fields and to be deformed with a two-armed pattern.
In this deformed disk, set of a two-armed ($m=2$) vertical p-mode oscillation and 
an axisymmetric ($m=0$) g-mode oscillation are considered.
They resonaltly interact through the disk deformation when their frequencies are the same.
This resonant interaction amplifies the set of the above oscillations in the case where 
these two oscillations have wave energies of oposite signs.
These oscillations are assumed to be excited most efficiently in the case 
where the radial group velocities of these two waves vanish at the same place.
The above set of oscillations is not unique, depending on the node number, $n$,
of oscillations in the vertical direction.
We consider that the basic two sets of oscillations correspond to the twin QPOs.
The frequencies of these oscillations depend on disk parameters such as strength of
magnetic fields.
For observational mass ranges of GRS 1915+105, GRO J1655-40, XTE J1550-564, and H1743-322,
spins of these sources are estimated.
High spins of these sources can be described if the disks have weak poloidal magnetic
fields as well as toroidal magnetic fields of moderate strength.
In this model the 3 : 2 frequency ratio of high-frequency QPOs is not related to their
excitation, but occurs by chance.  
\end{abstract}

\section{Introduction}

Many quasi-periodic oscillations (QPOs) have been observed in low-mass X-ray binaries (LMXBs).
Among them, high frequency QPOs are particularly interesting, since they
will be attributed to the innermost regions of relativistic accretion disks and thus clarification
of their origins will give important informations about disk structure in strong gravitational fields
as well as about mass and spin of the central sources.
High frequency QPOs in LMXBs are classified into two classes:
High frequency QPOs (HF QPOs) in black-hole LMXBs and kHz QPOs in neutron-star LMXBs
(van der Klis 2004).
HF QPOs of black-hole LMXBs have no time variation and in some sources they are observed in pairs 
whose frequency ratio is close to 3 : 2.
KHz QPOs in neutron-star LMXBs are also observed in some sources in pairs, but their frequency
ratio is not 3 : 2, but changes with correlated time variations. 
It is unclear whether both types of QPOs come from a common origin.

So far, many models have been proposed to describe HF QPOs of black-hole LMXBs, especially focusing on 
the 3 : 2 frequency ratio.
Typical models will be the relativistic precession model proposed in a series of papers by Stella \&
Vietri (1998) and Morisink \& Stella (1999), and the epicyclic resonant model
proposed by Klu\'{z}niak \& Abramowicz (2001) and Abramowicz \& Kluzniak (2001) and extensively studied
by their collaborators (e.g., Abramowicz et al. 2003a, b; Bursa et al. 2004; Rebusco 2004, Kluz\'{n}iak et al. 
2004, Hor\'{a}k 2008, Stuchlik et al. 2008a, b, Hor\'{a}k et al 2009 and others).
In a context different from the epicyclic resonant model, importance of a resonance 
on wave excitation has been emphasized, and 
the warp resonant model has been proposed (Kato 2003, 2004, 2008, Kato \& Fukue 2006, 
Ferreira \& Ogilvie 2008, Oktariani et al. 2010).
This model is based on resonant excitation of a set of positive-energy and negative-energy oscillations 
by interaction through disk deformation (Kato et al. 2011, Kato 2011b).
In addition to the above models, many discoseismic modes have been studied in order to examine their 
applicability to QPOs since Kato \& Fukue (1980) (see Wagoner 1999, Kato 2001 and Kato et al. 2008 for
reviews, and Wagoner 2012 for recent work).

In spite of many efforts by the above-mentioned work and by other reseraches, 
the twin QPOs observed in microquasars are not consistently described by one unique model (T\"{o}r\"{o}k et al. 2011).
This difficulty becomes prominent by recent findings that microquasars with twin QPOs have 
extremely high spins (McClintock et al. 2011).

In the case of kHz QPOs in neutron-star LMXBs, frequencies of twin QPOs and their time variation seem
to be well described by assuming that kHz QPOs are two-armed vertical p-mode oscillations trapped in 
the innermost region of disks (Kato 2011a,c,; 2012a,b,c).
In the case of twin HF QPOs of black-hole LMXBs, however, a direct application of trapped vertical 
p-mode oscillations seems not to well describe observations.

One of important clues to explore the origins of the twin HF QPOs of black-hole LMXBs will be why they 
appear only in or near to the very high state (steep power-law state) (Remillard 2005), where the disk
consists of both a power-law component and a thermal disk component.
In the present paper we assume that in the very high state the innermost disk is deformed from an axisymmetric
one to a two-armed one,\footnote{
In the warp resonant model (Kato 2004, 2008; Kato \& Fukue 2006) the disk deformation was assumed to be one-armed.
}
although the origin of the deformation is not explored.\footnote{
There are numerical simulations showing a deformation of the innermost part of disks.
Machida \& Matsumoto (2008) show that in a cool state of magnetized accretion disks,  
a torus is created in the innermost part of disks and deformed into a crescent-like shape. 
}
In such deformed disks we impose a set of a two-armed ($m=2$) vertical p-mode oscillation 
with negative energy and an axisymmetric ($m=0$) g-mode oscillation with positive energy.
They have the same frequency.
We then examine in what cases the set of the oscillations are resonantly excited through the disk deformation
in order to apply the results to the HF QPOs in black-hole LMXBs.

As is shown below, we consider two sets of the oscillations.
The frequency ratio of these two set of oscillations derived by the present model 
is not always 3 : 2, depending on disk parameters.
In the present model, magnetic fields in disks are important parameters to specify the frequencies of 
resonantly interacting oscillations.
By adjusting the parameters in reasonable ranges, we examine whether we can describe 
the frequencies of twin HF QPOs of high-spin microquasars (GRS 1915+105, GRO J1655-40, 
XTE J1550-564, and H1743-322).
One of important parameters is strength of poloidal magnetic fields,
because the radial epicyclic oscillations are strongly modified by poloidal magnetic fields 
(Fu \& Lai 2009).

\section{Outline of Present Model of HF twin QPOs}
 
\subsection{Two-Armed-Deformed Disks}

As mentioned above, we assume that in the transition to the very high state the disk is deformed to a state
with a two-armed pattern.
The patern may rotate slowly in the azimuthal direction with frequency $\omega_{\rm D}$.
In this paper, for simplicity, the pattern is assumed just standing, $\omega_{\rm D}=0$.
(A generalization to a case of $\omega_{\rm D}\not= 0$ is simple.)
We further assume that the two-armed deformation consists of two components concerning the node number(s) 
in the vertical direction.
One is the deformation with one node in the vertical direction, i.e., $n_{\rm D}=1$, and the other
with $n_{\rm D}=2$.
A $n_{\rm D}=1$ deformation is the lowest mode of deformations which is asymmetric with respect to the equator.
In summary, the sets of frequency, $\omega_{\rm D}$, azimuthal wavenumber $m_{\rm D}$, and the vertical 
node number $n_{\rm D}$ of the deformation, ($\omega_{\rm D}$, $m_{\rm D}$, $n_{\rm D}$), 
are both (0, 2, 1) and (0,2,2).

\subsection{Two Types of Oscillations and Their Propagation Regions}

Let us consider a set of oscillations.
One is an axisymmetric ($m=0$) g-mode oscillation, and the other is a two-armed ($m=2$) vertical
p-mode oscillation (see, e.g., Kato 2001 and Kato et al. 2008 for classification of disk oscillations).
They are assumed to have the same frequency.
These two oscillations can resonantly interact each other through the two-armed disk deformation,
since the difference between the azimuthal wavenumbers is two and equal
to $m_{\rm D}$.\footnote{
For the resonant interaction to occur, in addition to this condition concerning the difference of
azimuthal wavenumbers, the difference of the vertical node numbers of the two oscillations must be
1 or 2.
This is related to the fact that $n_{\rm D}$ is taken to $n_{\rm D}=1$ and $n_{\rm D}=2$.
As will be discussed later, we consider axisymmetric g-mode oscillations with $n=1$ and two-armed 
vertical p-mode oscillations with $n=2$ and $n=3$.
Hence, this condition is satisfied.
}
This resonant interaction through the disk deformation can excite these two oscillations simultaneously
when wave energies of these oscillations have opposite sighs (Kato et al. 2011, Kato 2011b, see also
Kato 2008).

The axisymmetric g-mode oscillation has a positive energy because of the very fact of axisymmetry.
The two-armed vertical p-mode oscillation has two propagation regions in the radial direction;
inner and outer propagation regions.
Between them, there are evanescent region and the radius of corotation resonance.
It is generally known that in the inner propagation region waves has a negative energy, 
while those in the outer propagation region have a positive energy (e.g., Kato 2001).\footnote{
Let us consider an oscillation whose frequency is $\omega$ and the azimuthal wavenumber is $m$.
Then, the corotation resonance occurs at the radius where $\omega=m\Omega$.
One of the propagaion region of the oscillation is inside the corotation radius, and the oscillation 
in the region has a negative energy since $\omega-m\Omega<0$ there.
}
Hence, if the propagation region of the axisymmetric g-mode oscillation
and the inner propagation region of 
the two-armed vertical p-mode oscillation are spatially overlapped, we can expect 
simultaneous growth of these two oscillations with the same frequency by their resonant coupling
through the disk deformation.
The spatial overlapping of propagation region really occurs, as is shown below and in figures 1 and 2. 

Before discussing one more resonant condition which we adopt in this paper, we must notice the
propagation regions of the two-armed vertical p-mode oscillation and of the axisymmetric
g-mode oscillation, and their parameter dependences.
First, we consider the $m$-armed vertical p-mode oscillations.
At the boundary between the propagation and evanescent regions, the oscillations have long wavelength 
in the radial direction, and the relation between the frequency and boundary radius can be
obtained by considering purely vertical oscillations (e.g., Kato 2011a).
In the case of vertically isothermal disks with no magnetic fields, by solving the vertical eigen-value 
problem for purely vertical oscillations we find that the relation is given by 
\begin{equation}
         (\omega-m\Omega)^2=n\Omega^2_\bot 
\label{2.1}
\end{equation}
(cf., Okazaki et al. 1987),
where $\Omega(r)$ is the angular velocity of disk rotation at the radius $r$, 
$\Omega_\bot(r)$ is the vertical epicyclic frequency, and
$n$ is a positive integer specifying the number of node of oscillations in the vertical
direction, i.e., $n=1,2,...$.\footnote{
In the case of vertical p-mode oscillations, $n$ starts from 1.
The case of $n=0$ is the p-mode.
}

In the case of adiabatic oscillations in polytropic gases with polytropic index $N$, 
the radii where the vertical p-mode oscillations become purely vertical are
\begin{equation}
    (\omega-m\Omega)^2=\left\{\begin{array}{llr}
         \Omega_\bot^2           & (n=1) &     \\
         (1+\gamma)\Omega_\bot^2 & (n=2) &     \\
         3\gamma\Omega_\bot^2    & (n=3) &
         \end{array}
         \right.
\label{2.2}
\end{equation}
(Perez et al. 1997, Silbergleit et al. 2001, Kato 2005), where $\gamma=1+1/N$.
If the disk are vertically isothermal but terminated at a certain height, $z_{\rm s}$, by hot corona, and
subject to toroidal magnetic fields which are distributed in the vertical direction so that
$c_{\rm A}^2/c_{\rm s}^2$ [$c_{\rm s}(r)$ and $c_{\rm A}(r)$ being respectively the acoustic and 
Alfv\'{e}n speeds due to the toroidal component of magnetic fields]
is constant in the vertical direction, the radius where the vertical p-mode oscillations becomes 
purely vertical is determined by (Kato 2012b)
\begin{equation}
  (\omega-m\Omega)^2=\biggr(\frac{c_{\rm s}^2+c_{\rm A}^2}
       {c_{\rm s}^2+c_{\rm A}^2/2}K_{n,s}+1\biggr)\Omega_\bot^2,
\label{2.3}
\end{equation}
where $K_{n,s}$ is a value resulting from the eigenvalue problem of the purely
vertical oscillations and depending on the node number, $n$, in the vertical direction 
and the dimensionless height of truncation, $z_{\rm s}/H$, $H$ being the scale height of disks.
The value of $K_{\rm n,s}+1$ is given in table 1 of Kato (2012c). 
$K_{n,s}$ is $n-1$ in the limit of no truncation and larger than $n-1$ in general cases. 

As is shown above, the radii between the propagation and evanescent regions depends on the disk structure.
Here, we write the boundary, for simplicity, in the form
\begin{equation}
  (\omega-m\Omega)^2=\psi\Omega_\bot^2
\label{2.4}
\end{equation}
and $\psi$ is a parameter depending on disk structure.
In the case of two-armed ($m=2$) vertical p-mode oscillations, 
the boundary between the inner propagation and evanescent regions is thus given by
\begin{equation}
   \omega=2\Omega-\psi^{1/2}\Omega_\bot.
\label{2.5}
\end{equation}

\begin{figure}
\begin{center}
    \FigureFile(80mm,80mm){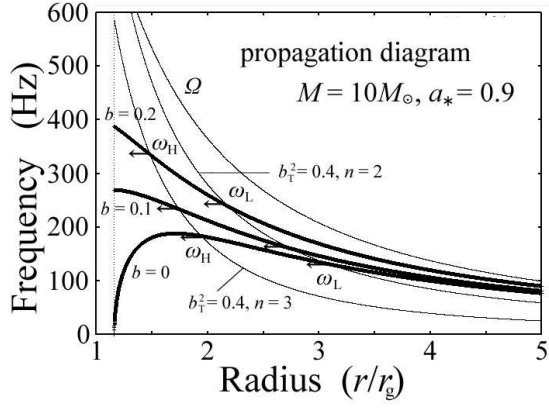}
\end{center}
\caption{Propagaton diagram (frequency - radius relation) for axisymmetric ($m=0$) g-mode
oscillations and two-armed ($m=2$) vertical p-mode oscillations in the case of $a_*=0.9$
and $M=10M_\odot$.
Three thick curves give the relation of $\omega={\tilde \kappa}$ for three cases of
$b=0$, 0.1, and 0.2 from bottom to top.
The propagation regions of axisymmetric g-mode oscillation in disks with $b=0$, 0.1, and 0.2
are, respectively, below the corresponding curve.
Among three thin curves the lower two curves are $\omega=2\Omega-(2.17)^{1/2}\Omega_\bot$
(i.e., $b_{\rm T}^2=0.4$ and $n=2$) and $\omega=2\Omega-(3.33)^{1/2}\Omega_\bot$ 
(i.e., $b_{\rm T}^2=0.4$ and $n=3$).
Two-armed vertical p-mode oscillations with $n=2$ and $n=3$ can propagate, respectively, 
below the corresponding curve in the case of $b_{\rm T}^2=0.4$.
The frequency given by the crossing point of two curves of $\omega=2\Omega-(2.17)^{1/2}\Omega_\bot$
and $\omega={\tilde \kappa}$ is the frequency of the lower HF QPOs, $\omega_{\rm L}$, in the present
model.
The waves with $\omega_{\rm L}$ propagate inside the point in the diagram (shown by arrow).
Similarly, the frequency given by the cross point of $\omega=2\Omega-(3.33)^{1/2}\Omega_\bot$
and $\omega={\tilde \kappa}$ is that of the upper HF QPOs.
The waves with $\omega_{\rm H}$ propagate inside the point in the diagram (shown by arrow).
 }
\end{figure}
\begin{figure}
\begin{center}
    \FigureFile(80mm,80mm){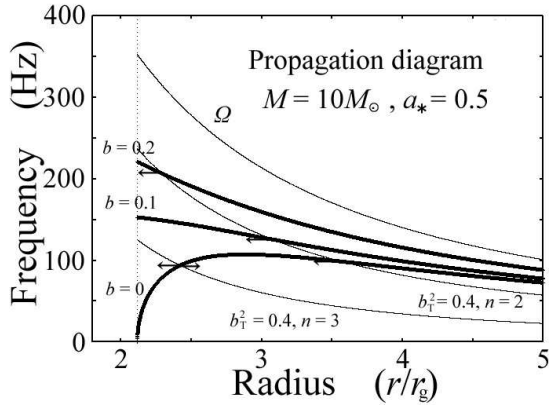}
\end{center}
\caption{The same as figure 1, except for $a_*=0.5$.
The curves of $\omega=2\Omega-(2.17)^{1/2}\Omega_\bot$ (i.e., $b_{\rm T}^2=0.4$ and $n=2$) do
cross with the curves of $\omega={\tilde \kappa}$ in three cases of $b=0$, 0.1, and 0.2.
The curve of $\omega=2\Omega-(3.33)^{1/2}\Omega_\bot$ (i.e., $b_{\rm T}^2=0.4$ and $n=3$) crosses 
with the curve of $\omega={\tilde \kappa}$ in the case of $b=0$, but not in other cases of $b=0.1$ and 0.2.
This shows the absence of $\omega_{\rm H}$ and HF QPO will be single.
It is noted that in the case of $b=0$, $\omega_{\rm H}$ is smaller than $\omega_{\rm L}$.
Hence, in such cases the naming of $\omega_{\rm L}$ and $\omega_{\rm H}$ due to the order of
frequency should be changed. 
}
\end{figure}

In general, $\psi$ is larger than $n$ [see equations (\ref{2.1}) -- (\ref{2.3})], and depends on 
$n$, $c_{\rm A}^2/c_{\rm s}^2$, and $z_{\rm s}/H$.
In order to avoid complexity, we consider hereafter vertically isothermal disks with no truncation,
i.e., $z_{\rm s}=\infty$, and $c_{\rm A}^2/c_{\rm s}^2$ is assumed to be radially constant.
Then, $\psi$ is a function of $b_{\rm T}^2$ for a given $n$ as [see equation (\ref{2.3})]
\begin{equation}
      \psi=\frac{1+b_{\rm T}^2}{1+b_{\rm T}^2/2}(n-1)+1,
\label{2.5'}
\end{equation}
where $b_{\rm T}^2$ is a parameter defined by $b_{\rm T}^2\equiv c_{\rm A}^2/c_{\rm s}^2$.
As is mentioned below, we are particularly interested in oscillations of $n=2$ and $n=3$.
In the case of $b_{\rm T}^2=0.4$, we have $\psi=2.17$ for $n=2$ and $\psi=3.33$ for $n=3$.
Hence, in order to demonstrate typical examples of the frequency - radius relations given by equation 
({\ref{2.5}) and (\ref{2.5'}), the relations in the case of $\psi=2.17$ and $\psi=3.33$ 
are shown in figures 1 and 2
by thin curves for different values of $a_*$.
Figure 1 is for $a_*=0.9$ and figure 2 is for $a_*=0.5$.
The mass of the central source is taken to be $10 M_\odot$ in both figures.
The left-hand side of the curves is the propagation region of the oscillations. 

Next, let us consider the propagation region of 
axisymmetric g-mode oscillations.
The region in the propagation diagram (frequency - radius diagram) is given by $\omega<\kappa$,
when there is no poloidal magnetic fields in disks, 
and the boundaries of the propagation region of axisymmetric g-mode oscillation is specified by
(Okazaki et al. 1987)
\begin{equation}
       \omega=\kappa.
\label{2.6}
\end{equation}
If poloidal magnetic fields exist, the radius specifying the boundary of propagation region of axisymmetric 
g-mode oscillations is strongly modified (Fu and Lai 2009).
In the case of axisymmetric g-mode oscillations with $n=1$, the propagation region is specified by
$\omega< {\tilde \kappa}$ and ${\tilde \kappa}$ is given by (Fu and Lai 2009)
\begin{equation}
    {\tilde \kappa}^2=\frac{1}{2}\biggr[\kappa^2+2\Omega_\bot^2b^2+(\kappa^4
                        +16\Omega_\bot^2\Omega^2b^2)^{1/2}\biggr],
\label{2.7}
\end{equation}
where $b\equiv {\tilde c}_{\rm A}/c_{\rm s}$, ${\tilde c}_{\rm A}$ being the Alfv\'{e}n speed defined by
the poloidal component of magnetic fields.
The frequency - radius relation given by 
\begin{equation}
            \omega={\tilde \kappa}
\label{2.8}
\end{equation}
is also shown in figures 1 and 2 by thick curves.
Three cases of $b=0$, 0.1, and 0.2 are shown in each figure.
The difference between figures 1 and 2 is the spin paramter, as mentiond before.

\subsection{A Working Hypothesis Determining Frequency of Resonant Oscillations}

The arguments given in subsection 2.2 are not enough to uniquely determine the frequency of  
resonantly excited oscillations, since the overlapping of the propagation regions is only a necessary condition.
We should consider what case the growth of oscillations occur most strongly.
To get a proper answer concerning this problem, detailed consideration on magnitude of coupling terms will be necessary.
Instead, we are satisfied here with adopting a simple working hypothesis, as was done in the case of 
the warp resonant model (Kato 2004, 2008).

The excitation due to the resonance will be non-steady in the present problem in the sense that a 
wave will be excited at a place where the resonant interaction  occurs most efficiently and 
then propagates in the radial direction.
This wave will not satisfy in general the trapping condition.
In other words, a possible condition determining the frequency of the excited oscillations 
will be that at the same radius the group velocities of both oscillations vanish together
so that the resonant interaction occurs efficiently.
In the limit of waves of short wavelength, the group velocity vanishes at the boundary between the propagation
and evanescent regions [see, for example, an expression for group velocity given by equation (12.16) by 
Kato et al. (2008)].
The boundaries are nothing but the radii of Lindblad resonances in the case of p- and g- mode oscillations.
In the above contexts, we impose, as the condition determining the frequency of resonant oscillation, that
the boundary radius between the propagation and evanescent regions is the same for two
resonantly interacting oscillations.
As mentioned before, the bounday between the propagation and evanescent regions is given by
equation (\ref{2.5}) for the two-armed vertical p-mode oscillations, and by
equation (\ref{2.8}) for the axisymmetric g-mode oscillation.
Thus, the crossing point of these two equations on the propagation diagram gives the frequency of
the excited oscillations in the present model (see figures 1 and 2).

In the followings, the above-mentioned condition is examined in two cases of a highly spinning source
($a_*=0.9$) and a less highly spinning source ($a_*=0.5$).
First, we consider the coupling between a two-armed vertical p-mode oscillation with $n=2$ and  
an axisymmetric g-mode oscillation with $n=1$, taking $a_*=0.9$ (figure 1).\footnote{
These two oscillations can resonantly interact through the disk deformation, since the difference of $n$'s is 1 
and is equal to one of $n_{\rm D}$.
}
The mass of the central source is taken to be $10M_\odot$.
If $\psi=2.17$ (i.e., $b_{\rm T}^2=0.4$ and $n=2$) and $b=0$ are adopted, the frequency is given 
by the crossing point of two curves of $\omega=2\Omega-(2.17)^{1/2}\Omega_\bot$ and $\omega=\kappa$ on the propagation diagram (see figure 1),
the point being one of those labelled by $\omega_{\rm L}$ in figure 1.
If $b=0.1$ is adopted with other parameters unchanged, the point moves in the left-upper direction in the figure.
If $b=0.2$ is adopted, the point moves further in the left-upper direction.
The point is also labelled by $\omega_{\rm L}$ in figure 1.
Similarly, we consider the interaction between the two-armed vertical p-mode oscillation with
$n=3$ and the axisymmetric oscillation with $n=1$.\footnote{
These two oscillations can resonantly interact since the difference of $n$'s is 2 and is
equal to one of $n_{\rm D}$.
}
Now, we take $\psi=3.33$ (i.e., $b_{\rm T}^2=0.4$ and $n=3$).
Then, the crossing point of two curves of $\omega=2\Omega-(3.33)^{1/2}\Omega_\bot$ and $\omega={\tilde \kappa}$
in three cases of $b=0$, 0,1, and 0.2 is shown in figure 1.
The points in two cases of $b=0$ and 0.2 are shown by labelling $\omega_{\rm H}$.
Here, we regards $\omega_{\rm L}$ and $\omega_{\rm H}$ in the same cases of parameters as the set of
frequencies of the lower and upper HF QPOs.

Next, we consider the case of $a_*=0.5$, keeping other parameters the same as those in corresponding 
cases of $a_*=0.9$.
As is shown in figure 2, we have no crossing point between the curve of  $\omega=2\Omega-(3.33)^{1/2}\Omega_\bot$ 
(i.e., $b_{\rm T}^2=0.4$ and $n=3$) 
and $\omega={\tilde \kappa}$ in the region outside the radius of the marginally stable circular orbit, except 
for the case of $b=0$.
This comes from the following situations.
Both curves of $\omega=2\Omega-\psi^{1/2}\Omega_\bot$ and $\omega={\tilde \kappa}$ shift downward on 
the propagation diagram as $a_*$ decreases, but the former much moves downward compared with the latter.
It is noted that in the case of no poloidal magnetic fields, the set of ($\omega_{\rm L}$, $\omega_{\rm H}$) 
always exist, but $\omega_{\rm L}> \omega_{\rm H}$.
That is, for small $a_*$ beyond at a certain value, the role of $\omega_{\rm L}$ and $\omega_{\rm H}$ is changed.

We should notice that in the present model, HF QPOs do not always appear in pairs as mentioned above.
That is, in the sources with high spin or weak poloidal magnetic fields, they can appear in pairs, but 
in other cases, there is the possibility of only one or no HF QPOs.

\section{Twin HF QPOs on Normalized Frequency -- Frequency Diagram}

In order to compare the set of $\omega_{\rm L}$ and $\omega_{\rm H}$ with observed frequencies of twin HF QPOs,
we plot the set of ($\omega_{\rm L}M$, $\omega_{\rm H}M$) on the $\omega_{\rm L}M$ - $\omega_{\rm H}M$
diagram, where $M$ is the mass of the central source.
The reason why we consider the $\omega_{\rm L}M$ -- $\omega_{\rm H}M$ diagram is
that the position of calculated ($\omega_{\rm L}M$, $\omega_{\rm H}M$) on the diagram is independent of $M$, 
and depends only on the parameters $a_*$, $b$ and $b_{\rm T}^2$.\footnote{
If $z_{\rm s}/H\not= \infty$, $z_{\rm s}/H$ is also an additional parameter.
}
For various sets of ($a_*$, $b_{\rm T}^2$), the position of ($\omega_{\rm L}M$, $\omega_{\rm H}M$) on the diagram
is shown for three cases of $b=0$, 0.1, and 0.2 (see subsequent subsections for individual source).
Figure 3 is for $b=0.1$, and figure 4 is for both $b=0$ and $b=0.2$.
The value of $a_*$ is changed continuously from $a_*=0.99$ towards $a_*=0$, while $b_{\rm T}^2$ is considered 
in four cases ($b_{\rm T}^2=0$, 0.2, 0.4, and 0.6) in figure 3 and in three cases
($b_{\rm T}^2=0$, 0.2, and 0.4) in figure 4. 
It is noted that the series of changing $a_*$ with $b=0.2$ are terminated at certain value of $a_*$
(depending on $b_{\rm T}^2$) before goint to $a_*=0$ (see the right-upper curves in figure 4).
This is due to the absence of $\omega_{\rm H}$ (and $\omega_{\rm L}$ too in some cases) as mentioned before
(compare figures 1 and 2). 

Before evaluating ($a_*$, $b$, $b_{\rm T}^2$) of four microquasars (GRS 1915+105, GRO J1655-40,
XTE J1550-564, and H1743-322) by using the figures, we first draw
a bird's eye view showing where twin HF QPOs of four microquasars are on this diagram.
To do so, we adopt the following data for the microquasars:
\begin{eqnarray}
  &&{\rm GRS\ 1915+105}\ \ \ :  \omega_{\rm L}^{\rm obs}=113{\rm Hz},\ \omega_{\rm H}^{\rm obs}=168{\rm Hz},\ M=10-18M_\odot  \nonumber \\
  &&{\rm GRO\ J1655-40}\ \ :  \omega_{\rm L}^{\rm obs}=300{\rm Hz},\ \omega_{\rm H}^{\rm obs}=450{\rm Hz},\ M=5.1-5.7M_\odot  \nonumber \\
  &&{\rm XTE\ J1550-564} \ :\omega_{\rm L}^{\rm obs}=184{\rm Hz},\ \omega_{\rm H}^{\rm obs}=276{\rm Hz},\ M=8.5-9.5M_\odot  \nonumber \\
  &&{\rm H1743-322}\ \ \ \ \ \ \ \ :\omega_{\rm L}^{\rm obs}=166{\rm Hz},\ \omega_{\rm H}^{\rm obs}=242{\rm Hz},\ M=5.0-15.0M_\odot.
\label{3.1}
\end{eqnarray}
Observed lower and upper frequencies of the twin HF QPOs are denoted by $\omega_{\rm L}^{\rm obs}$ and
$\omega_{\rm H}^{\rm obs}$ in order to distinguish them from the calculated ones.
Here, the observed mass ranges of GRS 1915+105, GRO J1655-40, XTE J1550-564, and H1743-322 are taken,
respectively, by referring to 
McClintock \& Remillard (2006), Beer \& Podsiadlowski (2002), Orosz et al. (2011), and Steiner et al. (2011),
respectively.

Because of uncertainty of mass, the position of ($\omega_{\rm L}^{\rm obs}M$, $\omega_{\rm H}^{\rm obs}M$)
on the diagram is on a finite line with a gradient of roughly 3:2 for each source.
Since the ranges of the lines for four sources overlap on the diagram, we present the ranges by introducing rectangles whose
diagonal shows the range.
Using this convention, we overdraw in figures 3 and 4 the ranges of position of 
($\omega_{\rm L}^{\rm obs}M$, $\omega_{\rm H}^{\rm obs}M$) of the observed HF QPOs of the four microquasars
by rectangles.

It is noted that the position of twin QPOs by the epicyclic resonant model (Ep) (Kluzniak \& Abramowicz 2001)
and the warp resonant model (Wp) (Kato 2008) are also shown, for reference, in figure 3 for some values of $a_*$.
These models have no other parameter than $a_*$.
Hence, the position of twin HF QPOs by these models is determined only by $a_*$.

In figure 3 (and figure 4) the most probable points of ($\omega_{\rm L}^{\rm obs}M$, $\omega_{\rm H}^{\rm obs}M$) 
of the four microquasars on the ($\omega_{\rm L}M$, $\omega_{\rm H}M$) diagram are all roughly around the center 
of the figure.
This suggests that the four microquasars belong all to a similar class.
Furthermore, comparison of figure 3 with figure 4 shows that $b=0.1$ is better than $b=0$ and $b=0.2$ to
describe observations.
That is, figure 4 shows that if $b=0$, the present model cannot describe observations even if $a_*$ is taken as high 
as $a_*\sim 1$, 
unless each mass of microquasars is taken to be smaller than each mean value in the observationally
allowed range given in equation (\ref{3.1}).
Figure 4 further shows that $b=0.2$ is too large to describe observations in the same sense mentioned above.
In other words, if the present model is correct, the poloidal magnetic fields in disks in these four
microquasars will roughly be around $b\simeq 0.1$.
   
Restricting only to the case of $b=0.1$, we estimate the values of other parameters, especially the spin parameter
$a_*$ in order to examine whether it is consistent with that estimated from spectral analyses.
Hereafter, we discuss the four microquasars separately.

\begin{figure}
\begin{center}
    \FigureFile(120mm,120mm){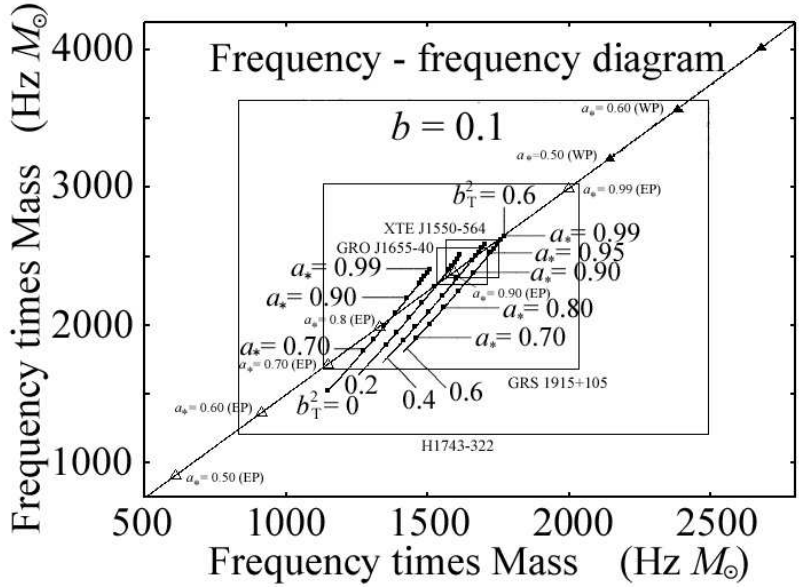}
\end{center}
\caption{
Relation between $\omega_{\rm L}M$ and $\omega_{\rm H}M$ on frequency - frequency diagram normalized by
mass of the central source, i.e., the coordinates are frequency times mass.
The relation is specified by three parameters, which are $a_*$, $b$, and $b_{\rm T}^2$.
The latter two are related to magnetic structure of disks, i.e., $b$ is to the strength of poloidal magnetic
fields and $b_{\rm T}^2$ to that of toroidal magnetic fields.
In this figure $b$ is taken to be $b=0.1$.
Four curves are for $b_{\rm T}^2=0.0$, 0.2, 0.4, and 0.6 (from left to right),
$a_*$ being changed from 0.99 to lower values.
The upper end of each curve is for $a_*=0.99$.
Positions of four microquasars (GRS 1915+105, GRO J1655-40, XTE J1550-564, H1743-322) on the diagram are
shown by rectangles. That is, due to uncertainties of mass, the position of ($\omega_{\rm L}M$, $\omega_{\rm H}M$)
of each source on the diagram is on the line of diagonal of each rectangle specified by each source.
For reference, the position of frequency - frequency relation  by epicyclic model (Ep) and warp model (Wp)
are also shown by open angles and filled angles, respecively.  
The straight line running from left-bottom to right-top is the line on which the frequency ratio is 3 : 2.
}
\end{figure}

\begin{figure}
\begin{center}
    \FigureFile(120mm,120mm){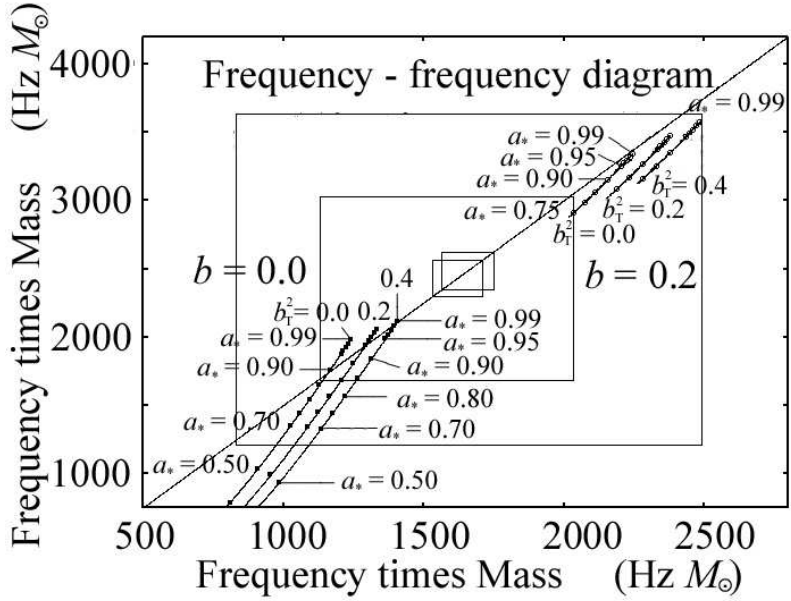}
\end{center}
\caption{The same as figure 3 except that in this figure the frequency - frequency relation in 
two cases of $b=0$ and $b=0.2$ are shown.
}
\end{figure}

\subsection{GRS 1915+105}

Figure 5 is an extension of a part of figure 3.
The diagonal of the large rectangle covering the alomst all area of figure 5 shows the range of
the possible position of the set of observed twin QPOs, when the mass range is taken to be $M=10-18M_\odot$.
The small filled rectangles on the diagonal in figure 5 is the position of the twin QPOs in the cases where 
the mass is taken from $10M_\odot$ to $18M_\odot$. 
The four lines running in the middle region of the figure are for $b_{\rm T}^2=0.0$, 0.2, 0.4, and
0.6 from left to right, changing the value of $a_*$.
On each line the upper end point is for $a_*=0.99$. 
This figure shows that the set of ($a_*$, $b_{\rm T}^2$) which can describe the observed set of QPO frequencies
are ($\sim 0.93$, $\sim 0.3$) if $M=14M_\odot$, and ($\>0.99$, $\sim 0.7$) if $M=16M_\odot$.
If $M>16M_\odot$, the present model cannot describe the observed twin QPOs as long as $b=0.1$ is adopted, 
since the value of $a_*$ required becomes larger than unity.
In other words, if this source has really $M>16M_\odot$ and the present model is correct, the poloidal
magnetic fields of this source are slightly stronger than $b=0.1$.
It is noticed that if $b$ is larger than $b=0.1$, the curves on figure 5 shift in the right-upward direction,
and observations can be described by $a_*$ smaller than unity (compare figures 3 and 4).
A large value of $a_*$ presented here is consistent with $a_*>0.98$ derived from
spectral analyses (McClintock \& Remillard 2006).

\begin{figure}
\begin{center}
    \FigureFile(80mm,80mm){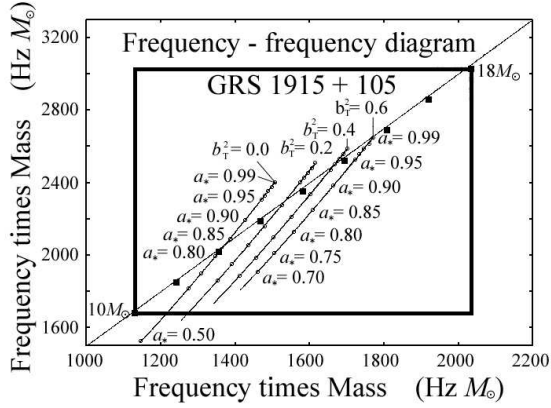}
\end{center}
\caption{An extension of figure 3 so that we can inspect what set of ($a_*$, $b_{\rm T}^2$) well describe  
the observed twin HF QPOs of GRS 1915+105 in the case where $b=0.1$ is adopted.
}
\end{figure}

\subsection {{\rm GRO J1655-40}}

Figure 6 is for GRO J1655-40, and the same as figure 5 except that the twin QPO frequencies are taken here to be 
$\omega_{\rm L}^{\rm obs}=300$Hz and $\omega_{\rm H}^{\rm obs}=450$Hz.
Figure 6 shows that the set of ($a_*$, $b_{\rm T}^2$) which can describe observations are ($\sim 0.90$, 0.2) 
when $M=5.1M_\odot$, and ($\sim 0.95$, $\sim 0.4$) when $M=5.5M_\odot$, and ($\sim 0.99$, $\sim 0.5$) 
when $M=5.7M_\odot$.
The spin estimated from spectral analyses is $a_*=0.70-0.80$ (Shafeee et al. 2006) and
$a_*\sim 0.98$ (Miller et al. 2011).
Our results are closer to the latter, as long as $b=0.1$ is adopted.
If $b$ is slightly larger than $b=0.1$, however, a smaller $a_*$ (with a smaller $b_{\rm T}^2$)
fits observations (compare figures 3 and 4).

\begin{figure}
\begin{center}
    \FigureFile(80mm,80mm){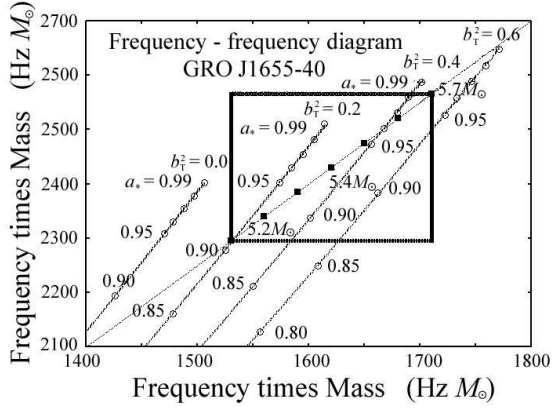}
\end{center}
\caption{An extension of figure 3 so that we can inspect what set of ($a_*$, $b_{\rm T}^2$) well describe  
the observed twin HF QPOs of GRO J1655-40 in the case where $b=0.1$ is adopted.
}
\end{figure}

\subsection{{\rm XTE J1550-564}}

Figure 7 is for XTE J1550-564, since $\omega_{\rm L}^{\rm obs}=184$Hz and $\omega_{\rm H}^{\rm obs}=276$ Hz 
are adopted.
This figure shows that the set of ($a_*$, $b_{\rm T}^2$) which can describe observed twin QPO frequencies
are ($\sim$0.92, $\sim 0.2$) when $M=8.5M_\odot$, and ($\sim 0.96$, 0.4) for $M=9.1M_\odot$.
The value of spin parameter 
$a_*$ required by the present model is rather high compared with the value derived from 
spectral analyses.
The latter is $a_*\sim 0.5$ in average (Steiner et al. 2011) or $a_*=0.70-0.75$ (Miller et al. 2009).

This difference will be accounted for if this source has a stronger poloidal magnetic fields than 
GRS 1915+105 in the sense that $b>0.1$.
This is because if $b>0.1$, the calculated curves in figure 7 shift in the upper-right direction
in the figure, as mentioned before in relation to GRO J1655- 40 (compare figures 3 and 4).

\begin{figure}
\begin{center}
    \FigureFile(80mm,80mm){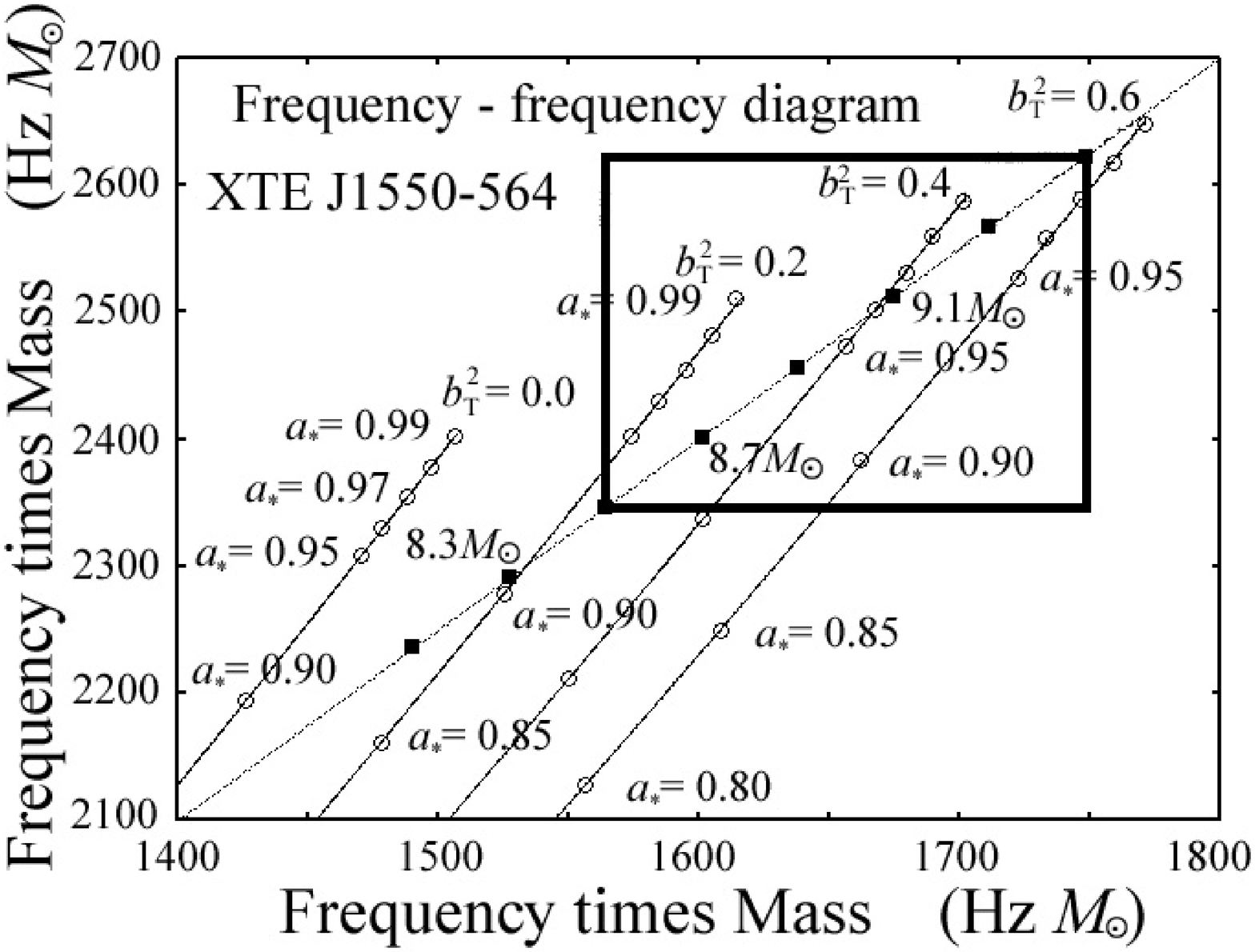}
\end{center}
\caption{An extension of figure 3 so that we can inspect what set of ($a_*$, $b_{\rm T}^2$) well describe  
the observed twin HF QPOs of XTE J1550-564 in the case where $b=0.1$ is adopted.
}
\end{figure}

\subsection{{\rm H1743-322}}

Figure 8 is for H1743, since this diagram is drawn by taking $\omega_{\rm L}^{\rm obs}=166$ Hz and 
$\omega_{\rm H}^{\rm obs}=$242 Hz.
This figure shows that the set of ($a_*$, $b_{\rm T}^2$) which can describe observations are
($\sim$0.80, $\sim$ 0) for $M=8M_\odot$, ($\sim$0.9, $\sim$0.2) for $M=9M_\odot$, and (0.93, 0.4) 
for $M=10M_\odot$.
The present model cannot describe observations if $M>11M_\odot$, as long as $b=0.1$ is adopted.
Steiner et al. (2011) estimate $a_*$ to be smaller than 0.7 by spectral analyses.
This difference of our $a_*$ and that of Steiner et al. (2011) can be accounted for if stronger 
poloidal magnetic fields than $b=0.1$ 
are present in H1743, as in the case of XTE J1550-564.

\begin{figure}
\begin{center}
    \FigureFile(80mm,80mm){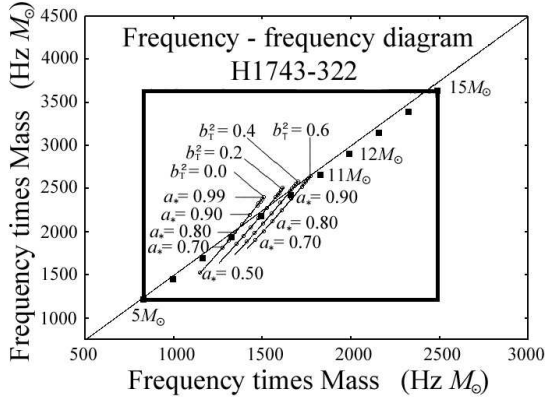}
\end{center}
\caption{An extension of figure 3 so that we can inspect what set of ($a_*$, $b_{\rm T}^2$) well describe  
the observed twin HF QPOs of H1743- 322 in the case where $b=0.1$ is adopted.

}
\end{figure}

\section{Discussion}

A standpoint of this model is that the 3 : 2 frequency ratio observed in twin HF QPOs of
microquasars does not come from their excitation processes, but occurs by chance.
In the present model, a quasi-periodic oscillation is a result of a resonant interaction
between two oscillations with positive and negative energies through disk deformation 
(Kato et al. 2011).
As the disk deformation, a two-armed ($m_{\rm D}=2$) deformation is assumed, different from our 
previous studies where an one-armed ($m_{\rm D}=1$) deformation is considered (e.g., Kato 2004, 2008).
As the set of positive and negative energy waves, we consider an axisymmetric ($m=0$) g-mode oscillation 
with positive energy and a two-armed ($m=2$) vertical p-mode oscillation with negative energy.

The difference of azimuthal wavenumber in the above two disk oscillations is two, which is 
equal to the azimuthal wavenumber, $m_{\rm D}$, of the disk deformation.
Hence, a necessary resonant condition among azimuthal wavenumbers of the two oscillations and 
the disk deformation is satisfied.
For the resonance really to occur, a relation among vertical node number is also necessary
for the two oscillations and the disk deformation.
In this paper we have considered {\it two} two-armed vertical p-mode oscillations with two nodes ($n=2$)
and three nodes ($n=3$).
On the other hand, as an axisymmetric g-mode oscillation, we have considered the mode with 
one node ($n=1$) in the vertical direction.
Hence, for these vertical p-mode oscillations to resonantly interact with
the g-mode oscillations through the disk deformation,
the disk deformation must have both components of $n=1$ and $n=2$.
The deformation with $n=1$ is the fundamental deformation of disks which is anti-symmetric
with respect to the equator.

One of vague points of this model is how much the assumption concerning the frequency of excited
oscillations (i.e., coincidence of boundaries of propagation regions for two interacting oscillations) 
is relevant or not.
To study this problem in details, we must examine functional forms in the radial and vertical directions
of two interacting positive and negative energy oscillations  and 
evaluate the volume integration of coupling terms by using those functional forms of oscillations and
disk deformation [see Kato (2008) as an example of detailed calculations of the coupling terms].
This is a complicated problem and beyond the scope of this paper.

Different from the kHz QPOs in neutron-star LMXBs, the observed frequencies of twin QPOs of black-hole 
LMBXs are robust, they being time-independent and their ratio being close to 3 : 2.
Furthermore, for four microquasars considered in this paper,
the observational points of ($\omega_{\rm H}^{\rm obs}M$, $\omega_{\rm L}^{\rm obs}$M) on the 
($\omega_{\rm H}$M, $\omega_{\rm L}M$) diagram are close when the typical observational values of $M$'s
are adopted for each source (see that the centers of four rectangles in figure 1 are close).
These observational evidences suggest that the disks of these sources are roughly in a similar state with 
similar time-independent values of disk parameters.
In the present model we have two dimensionless disk parameters describing the magnetic structure of the 
innermost region of disks.
They are $b$ and $b_{\rm T}$, where 
$b\equiv {\tilde c}_{\rm A}/c_{\rm s}$ and $b_{\rm T}\equiv c_{\rm A}/c_{\rm s}$,
${\tilde c}_{\rm A}$ and $c_{\rm A}$ being the Alfv\'{e}n speed defined by the poloidal and 
toroidal magnetic fields in disks, respectively, and $c_{\rm s}$ being the acoustic speed.
As discussed in the text, $b\geq 0.1$ and $b_{\rm T}=(0.2)^{1/2} \sim (0.4)^{1/2}=0.45 \sim 0.63$
seem to be relevant to describe observations.
Detailed comparison of the present model with observations will not be instructive at the present stage,
since the validity of the working hypothesis mentioned in subsection 2.3 might not be clear enough.
The strength of magnetic fields required to describe the QPO frequencies of the four microquasars 
by the present model is, however, in a narrow range as noticed above.
This may suggests that the strength of magnetic fields in disks is self-adjusted by a balance between winding by  
differential rotation and release by jet ejection.

In the present model, QPOs are not always in  pairs.
In some sources there is only one QPO or no QPO, depending on parameters.
For example, in disks with $b=0.2$ and $b_{\rm T}^2=0.4$, 
two QPOs ($n=2$ and $n=3$) are present when $a_*=0.9$ (see figure 1),  but 
QPO is only one ($n=2$) if $a_*$ is as small as 0.5 (see figure 2), because the
curve of $\omega=2\Omega-\psi^{1/2}\Omega_\bot$ for $n=3$ (with $b_{\rm T}^2=0.4$) is below
the curve of $\omega={\tilde \kappa}$ (with $b=0.2$) in all radius.
Even the QPO of $n=2$  disappears (there is no QPO) if $b_{\rm T}^2$ is slightly larger 
than $b_{\rm T}^2=0.4$ ($b=0.2$ and $a_*=0.5$), as understand from figure 2.
Next, let us consider sources with $b=0$.
In this case, there are always two QPOs as shown in figures 1 and 2.
However, in the case of figure 2 ($a_*=0.5$), the cross point between $\omega=\kappa$ and 
$\omega=2\Omega-\psi^{1/2}\Omega_\bot$ occurs inside the radius where $\kappa$ becomes the maximum,
as is shown in figure 2.
In this case the propagation regions of the two oscillations are in the oposite side of the crossing
point, and there is no overlapping region of the two oscillations.
In such cases, the coupling of the two oscillations and thus excitation of
the oscillations will be weak.
In summary, we can say that the sources where twin QPOs are observed are those where the 
disk parameters are in some limited range with strong spin parameter.
This might be one of reasons why twin QPOs are observed only in limitted sources. 

Finally, we should note that we have an important disk parameter which 
was not considered in this paper, but will affect results.
This is the vertical thickness of disks, $z_{\rm s}$.
In this paper we assumed that the disk is isothermal in the vertical direction and extends infinitely.
In actual situations, the disks will be truncated in  the vertical thickness at  a certain height, $z_{\rm s}$,
by the presence of corona.
In such disks, the propagation region of the two-armed vertical p-mode oscillations is modified (Kato 2012b,c),
which changes $\omega_{\rm H}$ and $\omega_{\rm L}$.
This will be a problem to be examined.
It is noticed that retreat of the inner edge of disks from the radius of the marginally stable circular orbit
has no serious effects on results of the present model, since the oscillations treated in the model
are not standing ones and no inner boundary condition is imposed to the oscillations.  
 
\bigskip\noindent
A note added on July 24, 2012.

In order to examine validity of the wave excitation mechanism adopted in this paper
[i.e., excitation of a set of waves with positive and negative wave energies through a resonant coupling
with disk deformation (Kato et al. 2011)], we consider waves in tidally deformed disks of binary stars.
It is found that the tidal instability and superhump phenomena in dwarf novae can well be interpreted by
the mechanism.
This will be shown in a subsequent paper.

\bigskip
\leftskip=20pt
\parindent=-20pt
\par
{\bf References}
\par
Abramowicz, M.A.\& Klu\'{z}niak, W. 2001, A\&A, 374, L19\par
Abramowicz, M.A., Bulik, T., Bursa, M., \& Klu\'{z}niak, W. 2003a, A\&A, 404, L21\par
Abramowicz, M.A., Karas, V., Klu\'{z}niak, W., Lee, W.H., \& Rebusco, P. 2003b, PASJ, 55, 466\par
Beer, M.E., \& Podsiadlowski, P. 2002, MNRAS, 331, 351\par
Bursa, M., Abramowicz, M.A., Karas, V.W., \& Klu\'{z}niak, W. 2004, ApJ, 617, L45\par
Greene, J., Bailyn, C.D., \& Orosz, J.A. 2001, ApJ, 554, 1290\par
Ferreira, B.T., \& Ogilvie, G.I. 2008, MNRAS, 386, 2297\par
Fu, W. \& Lai, D., 2009, ApJ, 690, 1386\par
Harlaftis, E.T. \& Greiner, J. 2004, A\&A, 414, 13\par
Hor\'{a}k, J. 2008, A\&A, 486, 1\par
Hor\'{a}k, J., Abramowicz, M.A., Klu\'{z}niak, W., Rebsuco, P., \& Torok, G. 2009, A\&A, 499, 540\par
Kato, S. 2001, PASJ, 53, 1\par 
Kato, S. 2003, PASJ, 55, 801\par
Kato, S. 2004, PASJ, 56, 905\par
Kato, S. 2005, PASJ, 57, 699 \par
Kato, S. 2008, PASJ, 60, 111\par
Kato, S. 2010, PASJ, 62, 635 \par
Kato, S. 2011a, PASJ, 63, 125 (toriodal mag. fields)\par
Kato, S. 2011b, PASJ, 63, 617 (double diffusion) \par
Kato, S. 2011c, PASJ, 63, 861 (toroidal, kHz QPOs) \par
Kato, S. 2012a, PASJ, 64, in press \par
Kato, S. 2012b, PASJ, 64, in press \par
Kato, S. 2012c, PASJ, submitted \par
Kato, S. \& Fukue, J. 1980, PASJ, 32, 377\par
Kato, S. \& Fukue, J. 2006, PASJ, 58, 909\par
Kato, S., Fukue, J., \& Mineshige, S. 2008, Black-Hole Accretion Disks --- Towards a New paradigm --- 
  (Kyoto: Kyoto University Press), chaps. 3 and 11 \par
Kato, S., Okazaki, A.T., \& Oktariani, F. 2011, 63, 363\par
Klu\'{z}niak, W. \& Abramowicz, M.A. 2001, Acta Phys. Pol.B32, 3605\par
Kluz\'{n}iak, W., Abramowicz, M.A., Kato, S., Lee, W.H., \& Stergioulas, N. 2004, ApJ, 603, 89 \par  
Machida, M., \& Matsumoto, R. 2008, PASJ 60, 613\par
McClintock, J.E., Remillard, R. 2006, ARA\&A, 44, 49\par  \par
McClintock, J.E., Shafee, R., Narayan, R., Remillard, R.A., Davis, S.W., \& Li, L.
   2006, ApJ, 652, 518 \par
McClintock, J.E., Narayan, R., Davis, S.W., Gou, L., Kulkarni, A., Orosz, J.A., Penna, R.F., Remillard, R., \& Steiner, J.F., 
   2011, Classical and Quantum Gravity, special volume for GR19, eds., D.Maraff, D. Sudarsky \par
Miller, J.M., Reynolds, C.S., Fabian, A.C., Miniutti, G., \& Gallo, L.C. 2009, ApJ, 697,900\par
Morsink, S.M. \& Stella, L. 1999, ApJ, 513, 827\par
Okazaki, A.T., Kato, S., \& Fukue, J. 1987, PASJ, 39, 457\par
Oktariani, f., Okazaki, A.T., \& Kato, S. 2010, 62, 709\par 
Orosz, J.A. et al. ApJ, 568, 8450 \par
Penninx, W. 1989, in J.Hunt, \& B.Battrick, eds. "Proceedings of 
    the 23rd ESLAB symposium on Two Topics in X-ray Astronomy", Bologna, Sept. 1989,
    ESA publications ESA SP-296, 185 \par
Perez, C.A., Silbergleit, A.S., Wagoner, R.V., \& Lehr, D.E. 1997, ApJ, 476, 589 \par
Priedhorsky, W., Hasinger, G., \& Lawin, W.H.G., et al. 1986, ApJ, 306,L91 \par
Rebsuco, P. 2004, PASJ, 56,553\par
Remillard, R. 2005, Astron. Nachr. 326, 804 \par 
Shafee, R., McClintock, J.E., Narayan, R., Davis, S.W., Li, L.,\& Remillard, R.A. 2006, 
    APJ, 636, L113\par
Silbergleit, A.S., Wagoner, R.V., \& Ortega-Dodr\'{i}gues, M. 2001, ApJ., 548, 335\par
Steiner, J.F., Reis, R.C., McClintock, J.E., Narayan, R., Remillard, R.A., Orosz, J.A., Gou, L.,
    Fabian, A.C., \& Torres, M.A.P. 2002, MNRAS, 416, 941 \par
Stella, L., \& Vietri, M. 1998, ApJL, 492, L59 \par
Stuchlik, Z., Konar, S., Miller, J.C., \& Hledik, S. 2008a, A\&A, 489, 963\par
Stuchlik, Z., Kotrlova, A., \& Torok, G. AcA, 58, 441\par
T\"{o}r\"{o}k, G. 2011, A\&A, 531, 59 \par
T\"{o}r\"{o}k, G., Kotrlova, A., Sramkova, E., \& Stuchlik, Z. 2011, A\&A, 531, 59 \par
van der Klis, M. 2004, in Compact stellar X-ray sources (Cambridge University Press), 
   eds. W.H.G. Lewin and M. van der Klis (astro-ph/0410551)    \par
Wagoner, R.V. 1999, Phys. Rep. 311, 259\par 
Wagoner, R.V. 2011, ApJL, in press (ArXiv:1205.1783)\par
\bigskip\bigskip

\end{document}